\documentclass[10pt,letterpaper,twocolumn]{article} 

\usepackage{ol2}
\usepackage[draft,implicit=false]{hyperref}
\usepackage{amsmath}

\begin{document}

\twocolumn[ 

\title{Inverted-wedge silica resonators for controlled and stable coupling}


\author{Fang Bo,$^{1,2}$ Steven He Huang,$^1$ \c{S}ahin Kaya \"{O}zdemir,$^{1,*}$ \\Guoquan Zhang,$^{2}$ Jingjun Xu,$^{2}$ and Lan Yang$^{1,*}$}

\address{
$^1$Electrical and Systems Engineering, Washington University in St. Louis, St. Louis, MO 63130, USA
\\
$^2$The MOE Key Laboratory of Weak Light Nonlinear Photonics, TEDA Applied Physics Institute and School of Physics, Nankai University, Tianjin 300457, China
\\
$^*$Corresponding author: yang@ese.wustl.edu, ozdemir@ese.wustl.edu
}

\begin{abstract}
Silica microresonators with an inverted-wedge shape were fabricated using conventional semiconductor fabrication methods. The measured quality factors of the resonators were greater than $10^6$ in 1550 nm band. Controllable coupling from under coupling, through critical coupling, to over coupling was demonstrated by horizontally moving a fiber taper while in touch with the top surface of the resonator. Thin outer ring of the resonator provided a support for the fiber taper leading to a robust stable coupling.
\end{abstract}


 ] 

Whispering-gallery-mode (WGM) microresonators with high quality factors $Q$ are of importance in studying fundamental physics and in developing practical applications, including quantum information, telecommunications, nonlinear optics, sensing, and cavity optomechanics~\cite{vahala_nature_2003, matsko_jqe_2006, Ilchenko_jqe_2006, he_lpr_2013, aoki_nature_2006, xiao_oe_2008, zhu_natphoton_2010, he_natnano_2011, kim_apl_2010, dantham_nl_2013, kippenberg_science_2008}. This widespread use of WGM resonators stems from their ability to confine light within a small mode volume for extended durations of time via continuous total internal reflection which helps to enhance the light-matter interactions significantly.

Last decade has witnessed an increasing interest in the studies of WGM microresonators thereof WGM resonators of different geometries (e.g., sphere, disk, ring, toroid, bottle, etc) and materials (silica, silicon, silicon nitride, etc) have been proposed and fabricated~\cite{he_lpr_2013}; each having its own advantages and disadvantages over the others. For example, silica microtoroids~\cite{armani_nature_2003} have $Q$ of order of $10^8$ thanks to the significantly reduced scattering losses due to surface-tension-induced smooth surface as the result of a reflow process using $\rm{CO_2}$ laser~\cite{armani_nature_2003}. However, $\rm{CO_2}$ laser reflow process is not compatible with conventional semiconductor processing. Microdisk resonators, on the other hand, has lower $Q$ than microtoroids due to higher scattering losses. Recently, silica disk resonators with a wedge shape, referred to as wedge-resonator, has been fabricated by optimizing the fabrication process of microdisk resonators. Wedge resonators of millimeter sizes with $Q$ over $10^8$ have been reported~\cite{lee_natphoton_2012}. These resonators achieve such high-$Q$ values without the need for $\rm{CO_2}$ laser reflow, and due to their larger size they find uses in the studies of stimulated Brillouin scattering~\cite{lee_natphoton_2012} and optical combs~\cite{kippenberg_science_2011}.

Evanescent wave couplers such as prisms, waveguides or tapered fibers have been used to couple light in and out of WGM resonators. Taperoptimizinged fibers with coupling efficiencies over $95\%$ have been the most efficient scheme in almost all experiments with high-$Q$ resonators. However, maintaining a stable and accurately controlled coupling between the fiber taper and the resonator is challenging in the field outside the controlled laboratory environment. For example, airflow, mechanical vibrations and thermal fluctuations may easily perturb the coupling conditions and thus affect the transmission spectra of the resonator. Reflowed silica side walls or nanofork structures close to a resonator have been fabricated to form a support for the taper fiber coupler, providing mechanical stability and robust coupling conditions ~\cite{lin_prl_2009, monifi_jlt_2012}. In an alternative approach, resonators with different shapes, such as the one referred to as ``octagonal silica toroidal microcavity"~\cite{kato_apl_2012}, are fabricated to allow controllable coupling by changing the contact point between the resonator and the fiber taper. Since the fiber taper is in contact with the resonator, coupling is robust to external perturbations.

\begin{figure}
\begin{center}
\includegraphics[scale=0.48]{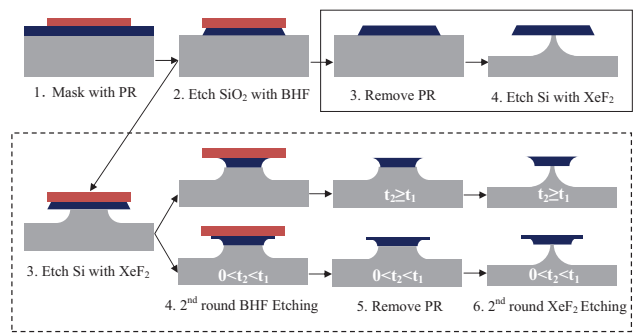}\\
\end{center}
\caption{Fabrication of wedge (solid box) and inverted-wedge (dashed box) shaped silica resonators. The brown, blue, and gray layers represent PR, silica, and silicon, respectively.}
\label{fig:fabrication_process}
\end{figure}

\begin{figure}
\begin{center}
\includegraphics[scale=0.5]{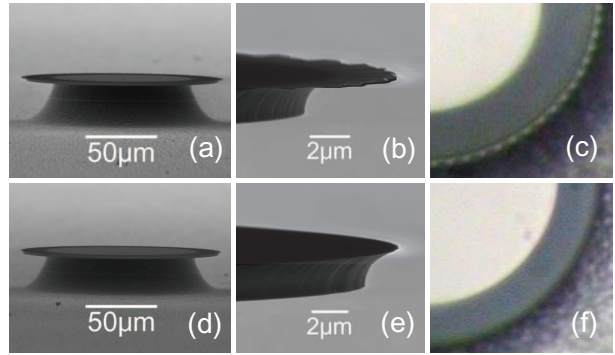}\\
\end{center}
\caption{Micrographs of the inverted-wedge shaped silica resonators. (a) and (d) are the SEM micrographs showing the side views of the resonators with and without a thin outer ring, whose edges are zoomed in and shown in (b) and (e), respectively. (c) and (f) present the optical micrographs depicting the top views of a quarter of the corresponding resonators, respectively.}
\label{fig:micrograph}
\end{figure}

In this work, we report a new type of high-$Q$ silica resonator fabricated by modifying the fabrication processes of microdisk and wedge resonators. We referred to this resonator as {\it inverted wedge resonator} because the top surface of this resonator is larger than the bottom surface in contrast to that of the wedge resonators~\cite{lee_natphoton_2012} (Figs.~\ref{fig:fabrication_process} and~\ref{fig:micrograph}). The resonator was fabricated using conventional semiconductor fabrication processes. Benefiting from its unique geometry with a flat top surface having a thin outer ring, robust critical coupling between tapered fiber and resonator was experimentally demonstrated with the outer ring serving as a support for the tapered fiber. The coupling strength was continuously tuned from under coupling to over coupling by scanning the position of the tapered fiber parallel to and in contact with the top surface of the resonator. This provides a very stable coupling between the taper and the resonator. Moreover, we also demonstrate vertical coupling between the resonator and the fiber taper by moving the fiber taper perpendicular to the top surface.

The fabrication process for the wedge resonators~\cite{lee_natphoton_2012} was modified to make the {\it inverted wedge} silica resonators. The detailed fabrication process is schematically illustrated in Fig.~\ref{fig:fabrication_process} as compared to that of the wedge-resonators. First, using photolithographic patterning circular photoresist (PR) pads were created on a silica-on-silicon wafer. Then, the substrate was immersed in buffered hydrofluoric (BHF) acid at room temperature for about $t_1\sim 20~{\rm min}$, and the exposed silica was isotropically etched downward and laterally, forming a wedge disk. This first BHF etching ensured that the exposed silica of thickness $2~\rm{\mu m}$  was just removed. With the PR mask intact, the wafer was put into $\rm{XeF_2}$ etcher which etched the exposed silicon isotropically for about $45~{\rm min}$, forming a silicon pillar under each silica disk. Then a second round of BHF etching was employed for a time duration of $t_2$, which etched the exposed silica laterally and vertically in upward direction as the bottom surface of the silica was exposed after the $\rm{XeF_2}$ etching, while the top surface was protected by the PR mask. For $0<t_2<t_1$, a thin outer ring was formed due to partial etching of the silica around the silicon pillar. The thickness of this silica layer can be controlled by adjusting $t_2$. When $t_2>t_1$ is chosen, the silica around the silicon pillar was removed completely. In both cases, the top surface of the microdisk is flat and larger than the bottom surface because BHF etches the silica from bottom to top. After this second BHF etching step, the PR was removed using acetone and a second round of $\rm{XeF_2}$ etching was performed. This last step decreased the size of the silicon pillar and further increase the separation between the silica wedge and the silicon substrate. Note that skipping the second BHF etching ($t_2=0$) prepares wedge silica resonators.

We fabricated a series of inverted-wedge resonators with diameters in the range of $135-138\rm{\mu m}$ by using different $t_2$. In Fig.~\ref{fig:micrograph}, we present the scanning electron microscope (SEM) micrographs together with the optical micrographs of typical resonators with and without the thin outer ring. These resonators were fabricated following the same recipe except for different $t_2$. The one in Fig.~\ref{fig:micrograph} (a-c) with a thin outer ring was obtained for
$t_2=19.5~{\rm min}$ whereas the one in Fig.~\ref{fig:micrograph} (d-f) without the outer ring was obtained for $t_2=20~{\rm min}$. From the side views of the fabricated resonators, it is clearly seen that these resonators differ from microdisk and wedge resonators, in the sense that the top surface is larger than the lower surface. Although the width of the outer ring is only 3 $\rm{\mu m}$, the stress-induced crown-like pattern reported in deep undercut resonators~\cite{chen_apl_2013} was observed in the optical micrograph (Fig.~\ref{fig:micrograph} (c)). This is attributed to the deformation of the very thin $\sim 50~\rm{nm}$ outer ring due to the induced stress as clearly seen in Fig.~\ref{fig:micrograph} (b).

\begin{figure}
\begin{center}
\includegraphics[scale=0.6]{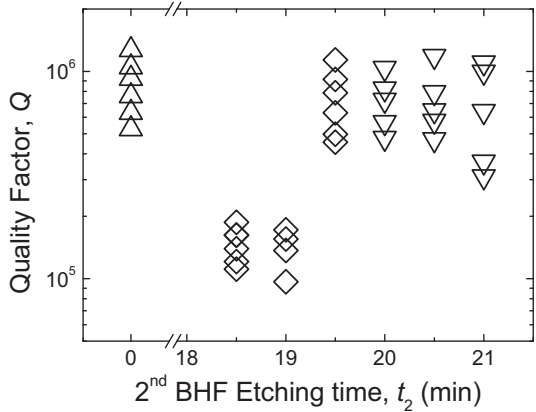}\\
\end{center}
\caption{Measured $Q$ of inverted-wedge resonators with different durations of the second round of buffered hydrofluoric acid  etching. The results for resonators with and without an outer ring are indicated by diamonds and inverted triangles, respectively.  As comparison, $Q$ of wedge-resonators fabricated through the same fabrication process but with $t_2=0$ are also shown by upright triangles.}
\label{fig:quality_factor}
\end{figure}

First we tested the effect of the duration of the second etching process $t_2$ on the intrinsic quality factor $Q_ {\rm{0}}$ of the resonators. Measurements were performed using light in the 1550 nm wavelength band from a tunable laser diode coupled to the resonators through a tapered fiber. The taper-resonator coupling was set to deep-undercoupling where the intrinsic quality factor equals to loaded quality factor (e.g., coupling losses are much smaller than the intrinsic losses). The input light power was set sufficiently low to prevent thermal effects on the transmission spectra while the wavelength of the laser was linearly scanned. Measured $Q$ for various resonators are given in Fig.~\ref{fig:quality_factor}.

As seen in Fig.~\ref{fig:quality_factor}, the highest $Q$ obtained for the resonators with thickness 2 $\rm{\mu m}$ and diameters  135-138 $\rm{\mu m}$  are $10^6$, which is on the same order as that of wedge-resonators of similar sizes~\cite{lee_natphoton_2012}. The highest $Q$ for resonators fabricated with and without the second BHF etching step are almost the same. This suggests that the second BHF etching doesn't deteriorate $Q$, but modifies the geometry of the resonator significantly. Shorter $t_2$ of the second BHF etching may decrease the $Q$ of inverted-wedge resonators due to mode leakage induced by a thicker outer ring. It is seen that resonators fabricated with $t_2=18.5~{\rm min}$ and $t_2=19.0~{\rm min}$ have $Q$ values which are almost an order of magnitude lower than those obtained for resonators fabricated with longer $t_2$. The $Q$ of these inverted-wedge  resonators should be able to reach $10^8$ as reported for wedge resonators~\cite{lee_natphoton_2012} by further optimizing the fabrication conditions, such as BHF concentration, silica wafer thickness, mask diameter, and etching times $t_1$ and $t_2$.

It is worth noting that the $Q$ of the resonator with a very thin outer ring, such as the resonators with $t_2=19.5~{\rm min}$ for the second BHF etching, is as high as those of the resonators without the outer ring. This is because the outer ring is too thin to support any optical mode. As a result, the optical mode is distributed in the thick part of the disk, with a similar distribution to those in resonators without the thin ring. Light cannot be coupled into the resonator if the taper is brought into contact with the outer part of the thin outer ring, where there are no mode overlap between the tapered fiber and the resonator. We experimentally confirmed this by comparing the power transmissions with the tapered fiber attached to and detached from the outer part of the resonator's outer ring. No observable transmission difference was measured at both on-resonance and off-resonance wavelengths. As the fiber taper was moved towards the center of the resonator while keeping it in contact with the resonator, the coupling strength gradually increased from under coupling to critical coupling and then to over coupling. Since the fiber taper was in contact with the surface of the resonator all the time, the coupling was very stable.

\begin{figure}
\begin{center}
\includegraphics[scale=0.6]{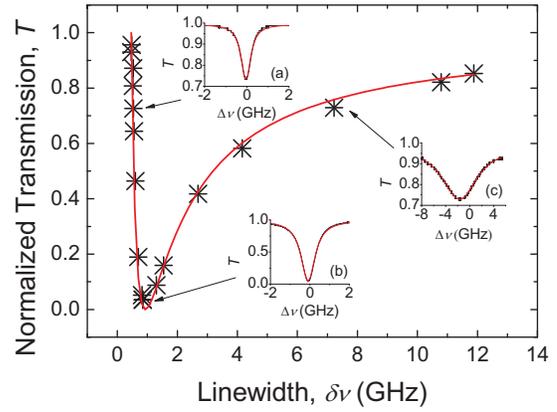}\\
\end{center}
\caption{Theoretical (red line) and measured normalized on-resonance transmission (black asterisk) as a function of the linewidth of the WGM for different horizontal positions of the tapered fiber. Insets (a), (b), and (c) show the typical transmission spectra (black curves) obtained for under coupling, critical coupling, and over coupling and their corresponding Lorentzian fitting curves (red curves), respectively.}
\label{fig:horizontal_control}
\end{figure}

Figure~\ref{fig:horizontal_control} shows the normalized transmission obtained in the experiments as a function of the linewidth at different positions of the fiber taper on the resonator. We controlled the coupling between the resonator and the fiber taper by moving the fiber taper on the surface of the inverted-wedge resonator with a thin outer ring. During this process, fiber taper was always in contact with the top surface of the resonator. We observed critical coupling point with an extinction of 19 dB  and over coupling with transmission larger than 80$\%$. The reason for the transmission to be less than unity in the overcoupling regime is the crosstalk between the mode of interest and the nearby lying modes. Theoretically, the relation between the normalized on-resonance transmission $T$ and the linewidth $\delta \nu$ follows
\begin{equation}
T=1-\frac{4}{Q_0}\left(\frac{\delta \nu}{\nu}-\frac{1}{Q_0}\right)\left(\frac{\nu}{\delta \nu}\right)^{2}.
\label{loading_curve}
\end{equation}
Using the measured $Q_{\rm{0}}=4.15\times 10^5$ and $\nu= 1.938\times10^{14}~\rm{Hz}$ as parameters, theoretical loading curve was plotted in Fig.~\ref{fig:horizontal_control} showing a good overlap with the experimental results. Typical normalized transmission $T$ spectra and their Lorentzian fitting curves in the under-, critical-, and over-coupling regimes are also shown in the insets (a), (b), and (c) of Fig.~\ref{fig:horizontal_control}, respectively. Although not shown here, we also tested vertical coupling between the fiber taper and the resonator by moving the fiber taper in the vertical direction at a fixed horizontal position on the resonator. The results are similar to those obtained for the horizontal case depicted in Fig.~\ref{fig:horizontal_control}. We obtained an extinction ratio of 20 dB at the critical coupling and a power transmission larger than 90$\%$ in the over coupling regime.

\begin{figure}
\begin{center}
\includegraphics[scale=0.6]{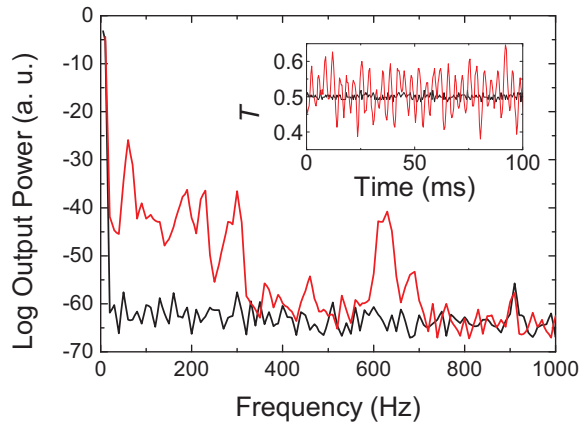}\\
\end{center}
\caption{Output spectra indicating the mechanical stability of the taper-resonator system. Inset shows the measured normalized transmission as a function of time. Black and red curves are when the taper is in contact and out of contact with the top surface of the inverted-wedge shaped resonator, respectively.}
\label{fig:stability}
\end{figure}

These inverted-wedge resonators allow a controllable and a stable coupling robust to external perturbations such as mechanical vibrations and air flow. We tested the stability of the coupling between the fiber and the resonator when the fiber taper is in contact and not in contact with the surface of the resonator. Note that when in contact, the outer ring served as a support plate for the tapered fiber, which fixed the distance and the mode overlap between the tapered fiber and the resonator, thereby stabilizing the coupling. In the last set of experiments, we set the taper-resonator coupling at the critical coupling and locked frequency of the input laser at around $T=1/2$ point. Under this condition, we monitored the transmission for several to tens of seconds. The results are depicted in Fig.~\ref{fig:stability} where we see that when the fiber taper is kept in contact with the top surface of the resonator, noise was significantly suppressed when compared to the case where the fiber taper was not in contact with the resonator. Typical experimentally obtained transmission values as a function of time are given in the inset of Fig.~\ref{fig:stability}, which clearly shows the suppression of fluctuations in the measured transmission when the taper is brought into contact with the top surface of the resonator.

In conclusion, inverted-wedge silica resonators with quality factors of the order of $10^6$ in 1550 nm communication band were fabricated by using only techniques compatible with conventional semiconductor processing. The coupling strength was tuned from under-coupling through critical coupling to over-coupling by scanning the relative position of the taper both horizontally and vertically with respect to the top surface of the resonator. Robust coupling between the tapered fiber and the resonator was achieved by placing the taper in contact with the top surface of the resonator. For in-field applications outside the laboratory settings, these resonators can be used without worrying about perturbations that can affect the coupling conditions. Moreover, if needed, the fiber taper and the resonator can be embedded in a polymer matrix~\cite{monifi_ieeeptl_2013} much easily than the previous packaging works without losing the coupling conditions when the polymer is injected, or the taper can be glued to the flat surface of the resonator in a way similar to spot-packaging~\cite{yan_ieeeptl_2011}. These resonators could be coupled to wedge resonators vertically or horizontally. Furthermore, since the top surfaces of these inverted-wedge resonators are untouched during the fabrication process, microstructures could be created on top of the silica wafer in advance.

This work was supported by the 973 programs (Grant Nos. 2011CB922003 and 2013CB328702), the NSFC (Grant Nos. 11374165
and 11174153), the 111 Project (Grant No. B07013), the China Scholarship Council, and ARO grant No. W911NF-12-1-0026. F. Bo thanks B. Peng, F. Monifi, X. Yang, J. Zhu, W. Chen, and Y. Xiao for their generous help and fruitful discussions.



\begin{thebibliography}{99}


\bibitem{vahala_nature_2003} K. J. Vahala,  Nature 424, 839 (2003).
\bibitem{matsko_jqe_2006} A. B. Matsko and V. S. Ilchenko,  IEEE J. Sel. Top. Quantum Electron. 12, 3 (2006).
\bibitem{Ilchenko_jqe_2006} V. S. Ilchenko and A. B. Matsko, IEEE J. Sel. Top. Quantum Electron. 12, 15 (2006).
\bibitem{he_lpr_2013} L. He, S. K. Ozdemir, and L. Yang,  Laser Photon. Rev. 7, 60 (2013).
\bibitem{aoki_nature_2006} T. Aoki, B. Dayan, E. Wilcut, W. P. Bowen, A. S. Parkins, T. J. Kippenberg, K. J. Vahala, and H. J. Kimble,
 Nature 443, 671 (2006).
\bibitem{xiao_oe_2008} Y.-F. Xiao, S. K. Ozdemir, V. Gaddam, C.-H. Dong, N. Imoto, and L. Yang, Opt. Express 16, 21462 (2008).
\bibitem{zhu_natphoton_2010} J. Zhu, S. K. Ozdemir, Y.-F. Xiao, L. Li, L. He, D.-R. Chen, and L. Yang,  Nat. Photon. 4, 46 (2010).
\bibitem{he_natnano_2011} L. He, S. K. Ozdemir, J. Zhu, W. Kim, and L. Yang,  Nat. Nano. 6, 428 (2011).
\bibitem{kim_apl_2010} W. Kim, S. K. Ozdemir, J. Zhu, L. He, and L. Yang,  Appl. Phys. Lett. 97, 071111 (2010).
\bibitem{dantham_nl_2013} V. R. Dantham, S. Holler, C. Barbre, D. Keng, V. Kolchenko, and S. Arnold,  Nano. Lett. 13, 3347  (2013).
\bibitem{kippenberg_science_2008} T. J. Kippenberg and K. J. Vahala, Science 321, 1172 (2008).
\bibitem{armani_nature_2003} D. K. Armani, T. J. Kippenberg, S. M. Spillane, and K. J. Vahala,  Nature 421, 925 (2003).
\bibitem{lee_natphoton_2012} H. Lee, T. Chen, J. Li, K. Y. Yang, S. Jeon, O. Painter, and K. J. Vahala,  Nat. Photon. 6, 369 (2012).
\bibitem{kippenberg_science_2011} T. J. Kippenberg, R. Holzwarth, and S. A. Diddams, Science 332, 555 (2011).
\bibitem{lin_prl_2009} Q. Lin, J. Rosenberg, X. Jiang, K. J. Vahala, and O. Painter, Phys. Rev. Lett. 103, 103601 (2009).
\bibitem{monifi_jlt_2012} F. Monifi, J. Friedlein, S. K. Ozdemir, and Y. Lan, J. Lightw. Technol. 30, 3306-3315(2012).
\bibitem{kato_apl_2012} T. Kato, W. Yoshiki, R. Suzuki, and T. Tanabe, Appl. Phys. Lett. 101, 121101 (2012).
\bibitem{chen_apl_2013} T. Chen, H. Lee, and K. J. Vahala,  Appl. Phys. Lett. 102, 031113 (2013).
\bibitem{monifi_ieeeptl_2013} F. Monifi, S. K. Odemir, J. Friedlein, and Y. Lan, IEEE Photon. Technol. Lett. 25, 1458 (2013).
\bibitem{yan_ieeeptl_2011} Y. Yan, C. Zou, S. Yan, F. Sun, J. Liu, C. Xue, Y. Zhang, L. Wang, W. Zhang, and J. Xiong, IEEE Photon. Technol. Lett. 23, 1736 (2011).

\end{thebibliography}
\end{document}